\documentclass[a4paper,10pt]{article}
\usepackage{pos}
\let\OLDthebibliography\thebibliography
\renewcommand\thebibliography[1]{
	\OLDthebibliography{#1}
	\setlength{\parskip}{0pt}
	\setlength{\itemsep}{0.00001pt}
}
\title{$|V_{cb}|$, LFU and $SU(3)_F$ symmetry breaking in $B_{(s)} \to D_{(s)}^{(*)} \ell \nu_\ell$ decays using Lattice QCD and Unitarity}
	
\author[a]{G.Martinelli}
\author*[b,c]{M. Naviglio}
\author[d]{S. Simula}
\author[e]{L. Vittorio}

\affiliation[a]{Physics Department and INFN Sezione di Roma La Sapienza, \\ Piazzale Aldo Moro 5, 00185 Rome, Italy}

\affiliation[b]{Dipartimento di Fisica dell’Università di Pisa, Largo Bruno Pontecorvo 3, I-56127 Pisa, Italy}

\affiliation[c]{Istituto Nazionale di Fisica Nucleare, Sezione di Pisa, Largo Bruno Pontecorvo 3, I-56127 Pisa, Italy}

\affiliation[d]{Istituto Nazionale di Fisica Nucleare, Sezione di Roma Tre,\\
Via della Vasca Navale 84, I-00146 Rome, Italy}

\affiliation[e]{LAPTh, Université Savoie Mont-Blanc and CNRS, Annecy, France}

\emailAdd{guido.martinelli@roma1.infn.it}
\emailAdd{manuel.naviglio@phd.unipi.it}
\emailAdd{silvano.simula@roma3.infn.it}
\emailAdd{ludovico.vittorio@lapth.cnrs.fr}

\abstract{We present an application of the unitarity-based dispersion matrix (DM) approach of Ref.~\cite{Unitarity} to the extraction of the CKM matrix element $|V_{cb}|$ from the experimental data on the exclusive semileptonic $B_{(s)} \to D_{(s)}^{(*)} \ell \nu_\ell$ decays~\cite{Martinelli:2021onb, Martinelli:2021myh, Bs_paper}. The DM method allows to achieve a non-perturbative, model-independent determination of the momentum dependence of the semileptonic form factors. Starting from lattice results available at large values of the 4-momentum transfer and implementing non-perturbative unitarity bounds~\cite{Martinelli:2021frl}, the behaviour of the form factors in their whole kinematical range is obtained without introducing any explicit parameterization of their momentum dependence. We consider the four exclusive semileptonic $B_{(s)} \to D_{(s)}^{(*)} \ell \nu_\ell$ decays and extract $|V_{cb}|$ from the experimental data for each transition~\cite{Martinelli:2021onb, Martinelli:2021myh, Bs_paper}. The average over the four channels is $|V_{cb}| = (41.2 \pm 0.8) \cdot 10^{-3} $, which is compatible with the latest inclusive determination at $1\sigma$ level. We address also the issue of Lepton Flavour Universality by computing pure theoretical estimates of the $\tau/\ell$ ratios of the branching fractions for each channel, where $\ell$ is a light lepton. In the case of a light spectator quark we obtain $R(D^*) =  0.275(8)$ and $R(D) = 0.296(8)$, which are compatible with the corresponding experimental values within $1.3\sigma$. In the case of a strange spectator quark we obtain $\textit{R}(D_s^*) =0.2497(60)$ and $\textit{R}(D_s) = 0.298(5)$. The different values for $R(D_s^*)$ and $R(D^*)$ may reflect $SU(3)_F$ symmetry breaking effects, which seem to be present in some of the lattice form factors, especially at large values of the recoil.}

\FullConference{%
  41st International Conference on High Energy physics - ICHEP2022\\
  6-13 July, 2022\\
  Bologna, Italy
}

\begin{document}
\maketitle

\section{State-of-the-art of exclusive semileptonic B-meson decays}
B decays are very challenging processes from a phenomenological point of view, principally because of two issues. The first one is the $|V_{cb}|$ puzzle, namely the observation of a tension~\cite{FLAG} between the exclusive and the inclusive determination of $|V_{cb}|$ at the level of $\simeq$ 2.8 standard deviations, namely
\begin{equation}
|V_{cb}|_{excl} \times 10^3 = 39.36(68) \ \ \ \ \ \ \ \ \ \ \ |V_{cb}| \times 10^3 = 42.00(65).
\end{equation}
Two new estimates of the inclusive value have also recently appeared, namely $|V_{cb}|_{incl} \times 10^3 = 42.16(50)$~\cite{Bordone} and $|V_{cb}|_{incl} \times 10^3 = 41.69(63)$ \cite{Bernlochner}, compatible with the inclusive FLAG value.

The second one is the discrepancy between the Standard Model predictions and experiments in the determinations of the $\tau/\ell$ ratios of the branching fractions, the so called $R(D^{(*)})$ anomalies, which represent an important test of Lepton Flavour Universality (LFU). They are defined as 
\begin{equation}
R(D^{(*)}) \equiv \frac{\Gamma(B \rightarrow D^{(*)}\tau \nu_{\tau})}{\Gamma(B \rightarrow D^{(*)}\ell \nu_{\tau})},
\end{equation}
where $\ell = e, \mu$ denotes a light lepton. The HFLAV Collaboration~\cite{HFLAV} recently computed the world averages of the available measurements of the $R(D^{(*)})$ ratios and of their SM theoretical expectations, obtaining
\begin{equation}
R(D)_{SM} = 0.298 \pm 0.004,  \ \ \ \ \ \ \ \ \ \ \ R(D)_{exp} = 0.339 \pm 0.026 \pm 0.014
\end{equation}
for the $B\rightarrow D$ case and 
\begin{equation}
R(D^*)_{SM} = 0.254 \pm 0.005, \ \ \ \ \ \ \ \ \ \ \ R(D^*)_{exp} = 0.295 \pm 0.010 \pm 0.010
\end{equation}
for the $B\rightarrow D^*$ one. As clearly stated by HFLAV Collaboration, the averages of the measurements of $R(D)$ and $R(D^*)$ exceed the corresponding SM predictions by $1.4\sigma$ and $2.8\sigma$, respectively. If the experimental correlation between these two quantities, namely $\rho = -0.38$, is taken into account, the resulting difference with the SM is increased to the $3.3\sigma$ level.

\section{The Dispersive Matrix Method}
Having these two issues, we need to investigate their nature more deeply. In this sense, it is fundamental to improve the precision with which we compute the form factors entering the hadronic matrix elements. Thus, we introduced a new method, the so called Dispersive Matrix (DM) method~\cite{Unitarity} based on an existing work~\cite{Lellouch}, whose main features we briefly recall here. Let us consider a generic form factor, $f(t)$, entering the hadronic matrix element of a generic $B \rightarrow Y^{(*)}\ell \nu_\ell$ decay, where $Y$ is generic a meson. Once the following quantities are given, namely 
\begin{enumerate}
\item The values of momentum transfer $t_1,...,t_N$ at which the form factor $f$ have been computed (e.g. on the lattice),
 \item the correspondent $f_1,...,f_N$ values of the form factors in that points,
 \item the susceptibility $\chi$ that are computed on the lattice~\cite{Unitarity,Martinelli:2021frl},
\end{enumerate}
then the properties of the method allow us to find bounds on the value of the form factor at a generic value of the momentum transfer. In particular, by defining $z_1, ..., z_N$ where $
z(t) = \frac{\sqrt{\frac{t_+-t}{t_+-t_-}}-1}{\sqrt{\frac{t_+-t}{t_+-t_-}}+1}$ with $t_{\pm} = (m_B \pm m_Y)^2$, the form factor in the point $z$ is bounded by unitarity, analyticity and crossing symmetry to be inside the interval
\begin{equation}
\beta(z) - \sqrt{\gamma(z)}\leq f(z) \leq  \beta(z) + \sqrt{\gamma(z)}
\end{equation}
where 
\begin{equation}
\beta(z) \equiv \frac{1}{\phi(z)d(z)}\sum_{j=1}^N \phi_j f_j d_j\frac{1-z^2_j}{z-z_j},  \ \ \ \ \ 
\gamma(z) \equiv \frac{1}{1-z^2}\frac{1}{\phi^2(z)d^2(z)}(\chi-\chi_{DM}),
\end{equation}
\begin{equation}
\chi_{DM} = \sum_{i,j=1}^N\phi_if_i\phi_jf_jd_id_j\frac{(1-z_i^2)(1-z_j^2)}{1-z_iz_j}.
\end{equation}
Here, $d(z)\equiv\prod_{m=1}^N(1-zz_m)/(z-z_m)$, $d_j \equiv\prod_{m\neq j=1}(1-z_jz_m)/(z_j-z_m)$ and the $\phi_j \equiv \phi(z_j)$ are the values of the kinematical function appropriate for the given form factor~\cite{Grinstein} containing the contribution of the resonances below the pair production threshold $t_+$. 
The obtained band of values represents the results of all possible BGL fits satisfying unitarity by construction and passing through the known points. The results do not rely on any assumption about the functional dependence of the form factors on the momentum transferred. Then, in this sense, they are model independent. Furthermore, the method is entirely based on first principles, the susceptibilities are non perturbative and we do not have series expansions. 

\begin{figure}[htb!]
\centering
 \includegraphics[scale=0.4]{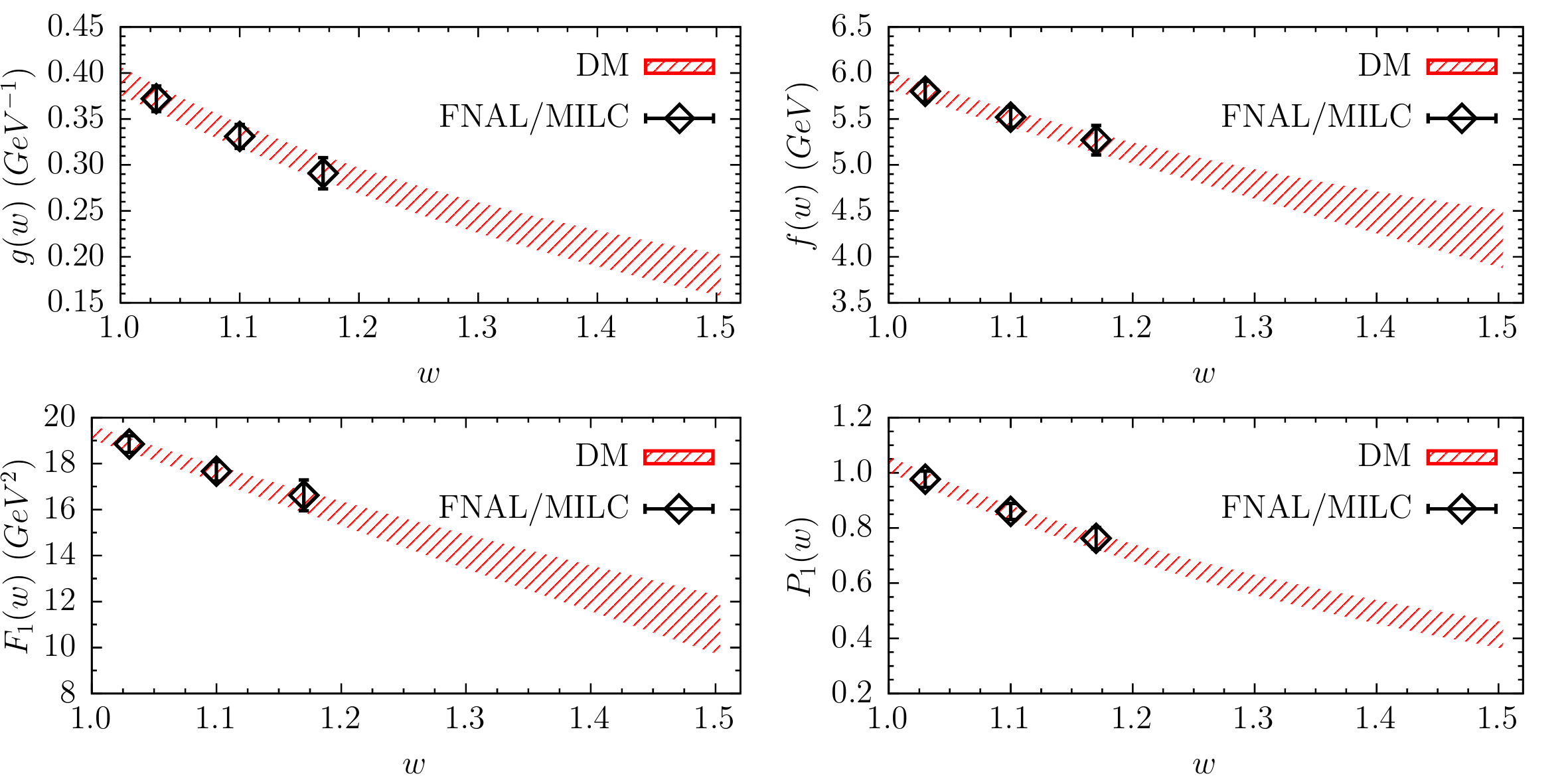}
 \centering
\caption{\textit{The bands of the FFs $g(w), f(w), F_1(w)$ and $P_1(w)$ computed by the DM method after imposing both the unitarity filter and the KCs. The FNAL/MILC values \cite{FNAL/MILC} used as inputs for the DM method are represented by the black diamonds.}\hspace*{\fill} \small}
\label{fig1}
\end{figure}

\section{The DM applied to the semileptonic $B \rightarrow D^{*}\ell \nu_\ell$ decays}

As an example that fully illustrates the effectiveness of the method, we discuss the case of the reconstruction of the form factors entering the semileptonic $B\rightarrow D^*$ matrix element~\cite{Martinelli:2021myh}. The Figure~\ref{fig1} shows the bands, covering the whole kinematic range, obtained using as inputs only the final results of the computations on the lattice performed by the FNAL/MILC Collaborations~\cite{FNAL/MILC}. There, in the ancillary files, the authors give the synthetic values of the FFs $g(w), f(w), \mathcal{F}_1(w)$ and $\mathcal{F}_2(w)$ at three non-zero values of the recoil variable $w$, namely $w=\{1.03,1.10,1.17\}$, together with their correlations. Note that the FF $\mathcal{F}_2$ is directly related to the $P_1(w)$ one, being $\mathcal{P}_1(w) = \mathcal{F}_2 \sqrt{r}/(1+r)$, where $r=m_{D^*}/m_B\simeq 0.38$. The DM unitarity bands are built up through bootstrap events that satisfy exactly both the unitarity filter and the Kinematical Constraints. These results show that the DM method allows to make predictions in the whole kinematical range with a quality comparable to the one obtained by the direct calculations, even if only a quite limited number of input lattice data are used. The bands are completely theoretical and come from a non-perturbative and model independent analysis, since no truncated z-expansion are present and no perturbative bounds are used. The method allows us to keep theoretical calculations and experimental data well separated in our analysis, since we do not want to introduce any bias that affects the shape of the form factors. 

\section{$|V_{cb}|$ and LFU Observables Results}
\begin{figure}[htb!]
\centering
 \includegraphics[scale=0.35]{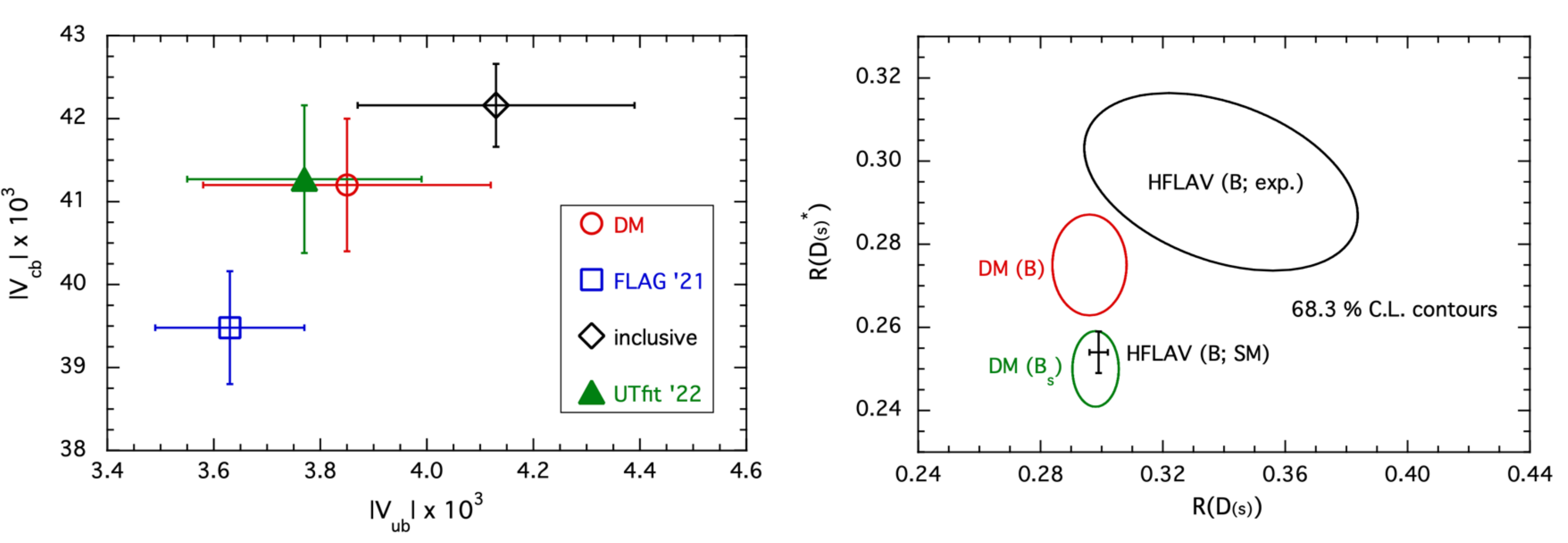}
 \centering
\caption{\textit{The left panel shows the DM values of $|V_{ub}|$ and $|V_{cb}|$ compared with the last results of the FLAG report~\cite{FLAG}, inclusive values \cite{Bordone,PDG} and UTfit indirect determinations \cite{UTFit1,UTFit2}. The right panel shows the pure theoretical estimates of the ratios $R(D^{(*)})$ compared with HFLAV averages from Ref. \cite{HFLAV}.}\hspace*{\fill} \small}
\label{fig5}
\end{figure}

The DM method has been adopted to compute the form factors of all semileptonic $B_{(s)}\rightarrow D_{(s)}^{(*)}$ decays~\cite{Martinelli:2021onb,Martinelli:2021myh,Bs_paper}. As in the $B \rightarrow D^*$ case, the method allows us to extract the relevant hadronic FFs in the whole kinematic range using only LQCD results available at large values of the 4-momentum transfer without making any assumption on their momentum dependence. The experimental data are never used to constraint the shape of the FFs but only to extract a determination of $|V_{cb}|$. This allows us also to extract pure theoretical estimates of $R(D)$ and $R(D^*)$. In Table~\ref{Vcb} we show the DM results for $|V_{cb}|$ for different channels and the correspondent average. As can be seen, for the first time there is an indication of a sizable reduction of the $|V_{cb}|$ puzzle.
In Table~\ref{Observables}, moreover we show our fully theoretical results for LFU and polarization observables. It can be observed also in this case that, for the first time, the $R(D^*)$ anomaly results to be lighter than the $2.5\sigma$ discrepancy stated by HFLAV~\cite{HFLAV}. Our findings are graphically collected in Fig.~\ref{fig5}, where it is also presented the DM estimate of $|V_{ub}|$~\cite{Pion, CKM}.

\begin{table}[t]
\caption[]{\textit{DM results for $|V_{cb}|$ for different channels. In the last row we show the corresponding average.}}
\label{Vcb}
\vspace{0.4cm}
\begin{center}
\begin{tabular}{|c|c|c|}
\hline
Process & Reference & $~|V_{cb}| \times 10^3~$\\
\hline
Inclusive $b\rightarrow c$  & Bordone et al., arXiv:2107.00604 & $~42.16 \pm 0.50~$\\
\hline
$B \rightarrow D$ & \textbf{DM method} & \textbf{~41.0 $\pm$ 1.2~}\\
& FLAG 2021, arXiv:2111.09849 & $~40.0 \pm 1.0~$\\
\hline
 $B \rightarrow D^*$ & \textbf{DM method} & \textbf{~41.3 $\pm$ 1.7~} \\
 & FLAG 2021, arXiv:2111.09849 & $~39.86\pm 0.88~$ \\
 \hline
 $B_s \rightarrow D_s$ & \textbf{DM method} & \textbf{~41.7 $\pm$ 1.9~ } \\
 \hline
 $B_s \rightarrow D_s^*$ & \textbf{DM method} & \textbf{~40.7 $\pm$ 2.4~} \\
& HPQCD Coll., arXiv:2105.11433 &  $~42.2 \pm 2.3~$\\
\hline 
 \ \  \textbf{Total Mean}    \   &   \ \ \ \ \ \ \ \ \ \ \ \ \  \textbf{DM method} \ \ \ \ \ \ \ \ \ \ \ \ \ &   \ \textbf{~41.2$\pm$ 0.8~ }      \ \\
 \hline
\end{tabular}
\end{center}
\end{table}
\begin{table}[htb!]
\caption[]{\textit{Fully theoretical results for LFU and polarization observables.}}
\label{Observables}
\vspace{0.4cm}
\begin{center}
{\small
\begin{tabular}{|c|c|c|c|}
\hline
Observable & \textbf{DM method} & Measurements & Difference \\
\hline
$R(D)$ &  \textbf{0.296(8)} & 0.340(27)(13) & $\simeq 1.3\sigma$\\
$R(D_s)$ &  \textbf{0.298(5)} & --- & ---\\
\hline
$R(D^*)$ & \textbf{0.275(8)} & 0.295(11)(8) & $\simeq 1.3 \sigma$\\
$R(D_s^*)$ & \textbf{0.2497(60)}   & --- & ---\\
\hline
$P_\tau(D^*)$ & \textbf{-0.529(7)} & $-0.38(^{+21}_{-16})$ & $<0.3\sigma$ \\
$P_\tau(D_s^*)$ & \textbf{-0.520(12)} & --- & ---\\
\hline
$F_L(D^*)$ & \textbf{0.414(12)} & 0.60(8)(4) & $\simeq 2.0 \sigma$\\
$F_L(D_s^*)$ & \textbf{0.440(16)} & --- & ---\\
\hline
\end{tabular}
}
\end{center}
\renewcommand{\arraystretch}{1.0}
\end{table}

\section{$SU(3)_F$ symmetry breaking effects}
\begin{figure}[htb!]
\centering
 \includegraphics[scale=0.35]{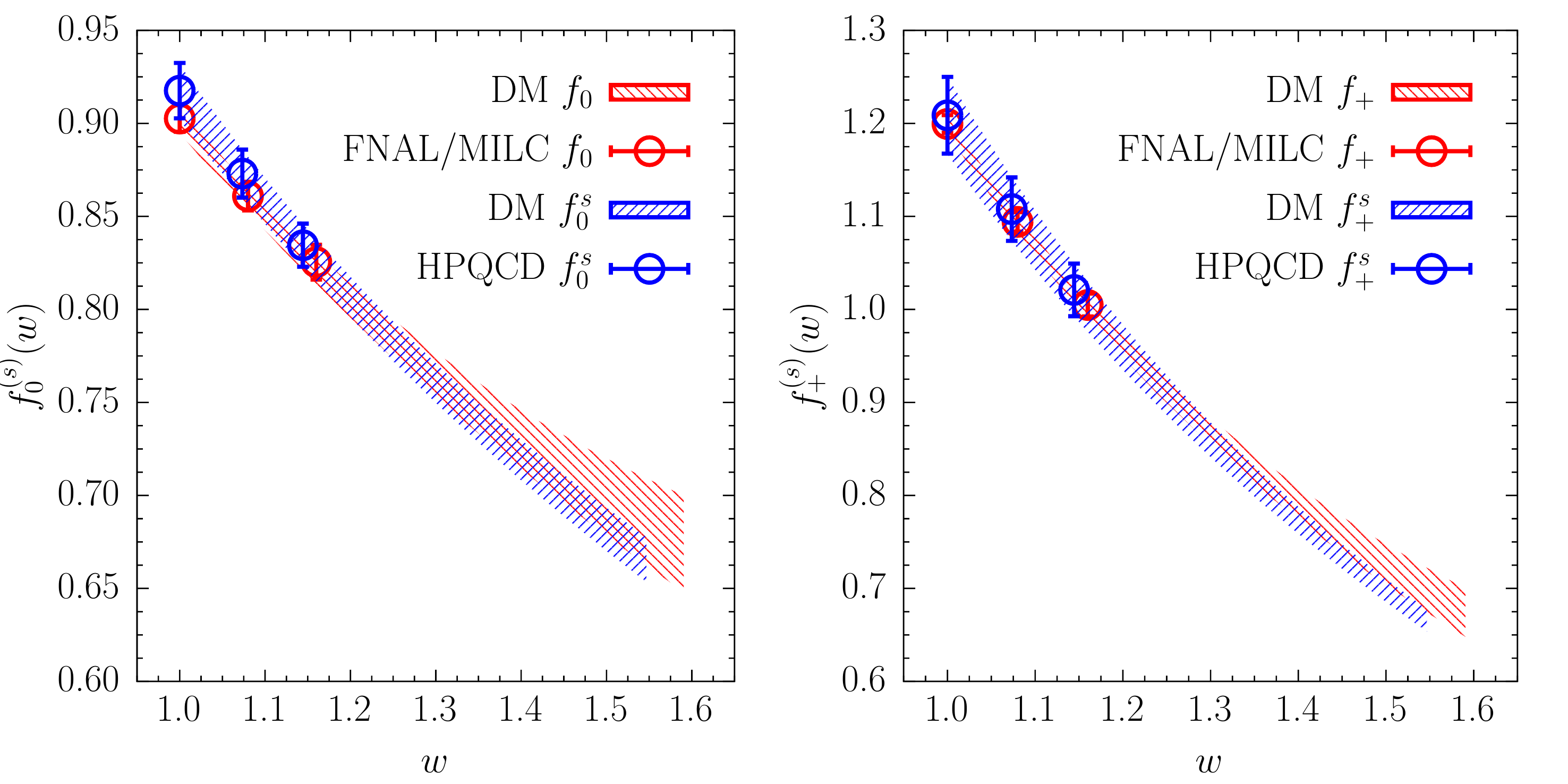}
 \centering
\caption{\textit{The DM bands of the scalar and vector FFs entering the semileptonic $B_{(s)} \rightarrow D_{(s)} \ell \nu_\ell$ decays versus the recoil variable w. The red and blue bands correspond to the $B\rightarrow D$ and $B_s \rightarrow D_s$ transitions, evaluated in Ref.~\cite{Martinelli:2021onb} and in Ref.~\cite{Bs_paper}. Correspondingly the red and blue circles represent the LQCD results used as inputs in the DM method, coming respectively from Refs.~\cite{MILC} and~\cite{McLean}.}\hspace*{\fill} \small}
\label{fig7}
\end{figure}

\begin{figure}[htb!]
\centering
 \includegraphics[scale=0.33]{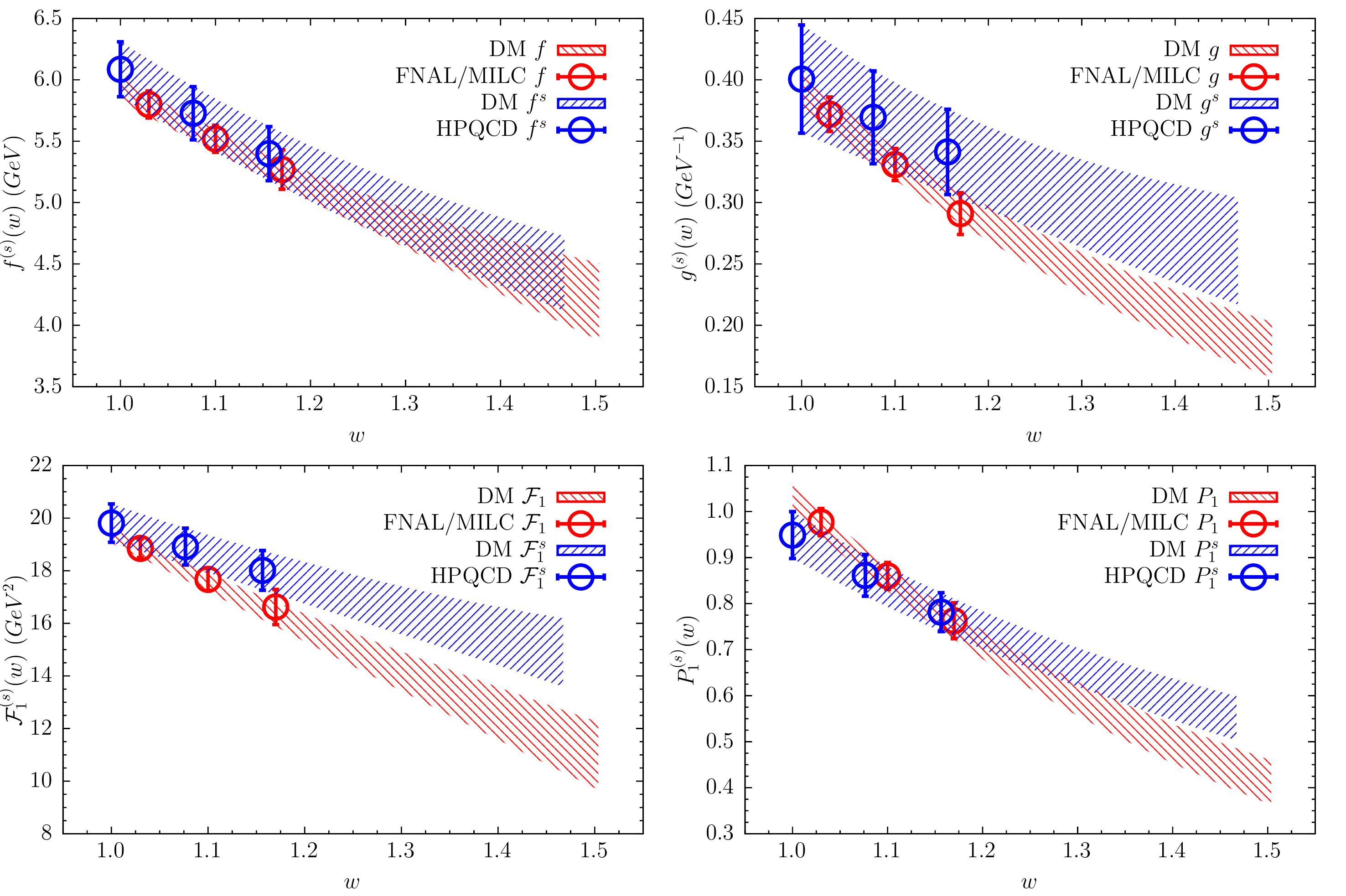}
 \centering
\caption{\textit{The DM bands of the four FFs entering the semileptonic $B_{(s)} \rightarrow D_{(s)}^* \ell \nu_\ell$ decays versus the recoil variable. The red and blue bands correspond to the $B\rightarrow D^*$ and $B_s \rightarrow D_s^*$ transitions, evaluated in Ref.~\cite{Martinelli:2021myh} and in Ref.~\cite{Bs_paper}, respectively. Correspondingly the red and blue circles represent the LQCD results used as inputs in the DM method, coming respectively from Refs.~\cite{FNAL/MILC} and~\cite{HPQCD}.}\hspace*{\fill} \small}
\label{fig8}
\end{figure}

At this point we want also to compare the DM bands of the hadronics FFs entering the $B\rightarrow D^{(*)}\ell\nu_\ell$, obtained in Refs.~\cite{Martinelli:2021onb,Martinelli:2021myh}, with the ones from $B_s\rightarrow D_s^{(*)}\ell \nu_\ell$, obtained in~\cite{Bs_paper}. The DM bands for the two sets of scalar and vector FFs of the  $B_{(s)}\rightarrow D_{(s)}\ell\nu_\ell$ transitions are shown in Fig~\ref{fig7}. The LQCD results used as inputs in the DM method come from Ref.~\cite{MILC} for the light spectator and from Ref.~\cite{McLean} for strange spectator. The bands turn out to be compatible with quite small differences in their slopes. Differently, the DM bands for the two sets of FFs of the $B_{(s)}\rightarrow D_{(s)}^*\ell\nu_\ell$ transitions are collected in Fig.~\ref{fig8}. The LQCD inputs used here come from Refs.~\cite{FNAL/MILC} and~\cite{HPQCD}, respectively. In this case the FFs $f$ and $g$ exhibit small $SU(3)_F$ breaking effects, while the shapes for the other two FFs are remarkably different. An interesting question is whether the spectator-quark dependence of the hadronic FFs shown in Figs.~\ref{fig8} is consistent with the experimental LHCb results on the ratios of branching fractions. The latter ones do not depend on $V_{cb}$ and are affected by  $SU(3)_F$ breaking effects within a $\simeq 10\%$ level. Using the DM bands of Figs.~\ref{fig8} we have evaluated the total decay rate modulo $|V_{cb}|^2$ for each of the four decay channels. Then, adopting the PDG values for the $B^0$ and $B_s$-meson lifetime, we obtain 
\begin{equation}
\frac{\mathcal{B}(B_s\rightarrow D_s \mu \nu)}{\mathcal{B}(B\rightarrow D\mu \nu)}\bigg|^{DM} = 1.02 \pm 0.06, \ \ \ \ \ \ \ \ \ \ \frac{\mathcal{B}(B_s\rightarrow D_s^* \mu \nu)}{\mathcal{B}(B\rightarrow D^*\mu \nu)}\bigg|^{DM} = 1.19 \pm 0.11.
\end{equation}
Within the present uncertainties of the order of $\approx$ 10 \% the above theoretical results agree with the corresponding experimental values in~\cite{LHCb} as well as with the updated results from Ref.~\cite{LHCb2}. A drastic improvement of the accuracy of both the experiments and the theory is mandatory in order to clarify the impact of  $SU(3)_F$ breaking effects on both the branching fractions and the hadronic FFs.


\end{document}